\documentstyle[IJFCS]{article}
\newtheorem{The}{Theorem}[section]
\newtheorem{Pro}[The]{Proposition}
\newtheorem{Deff}[The]{Definition}
\newtheorem{Lem}[The]{Lemma}
\newtheorem{Rem}[The]{Remark}

\newcommand{\fa}{\forall}
\newcommand{\Ga}{\Gamma}
\newcommand{\Gas}{\Gamma^\star}
\newcommand{\Gao}{\Gamma^\omega}
\newcommand{\Si}{\Sigma}

\newcommand{\Sio}{\Sigma^\omega}
\newcommand{\ra}{\rightarrow}
\newcommand{\hs}{\hspace{12mm}

\noi}

\newcommand{\ite}{\item}

\newcommand{\ol}{ $\omega$-language}

\newcommand{\om}{\omega}
\newcommand{\nl}{\newline}
\newcommand{\noi}{\noindent}

\newcommand{\de}{deterministic }

\begin{document}


\copyrightheading


\symbolfootnote

\textlineskip

\begin{center}

\fcstitle{ON RECOGNIZABLE LANGUAGES\\
        OF INFINITE PICTURES}

\vspace{24pt}

{\authorfont OLIVIER FINKEL \footnote{E Mail: finkel@logique.jussieu.fr}}

\vspace{2pt}

\smalllineskip

{\addressfont Equipe de Logique Math\'ematique   
 \\ U.F.R. de Math\'ematiques, Universit\'e Paris 7 \\  2 Place Jussieu 75251 Paris
 cedex 05, France}

\vspace{20pt}
\publisher{(received date)}
{(revised date)}{Editor's name}

\end{center}

\alphfootnote

\begin{abstract}
\noi  In a recent paper, Altenbernd, Thomas and 
W\"ohrle have considered acceptance of 
languages of infinite two-dimensional words (infinite pictures) by finite tiling systems,  
with the usual acceptance conditions, such as  the 
B\"uchi and Muller ones,  firstly  used for infinite words.
The authors asked for  comparing the tiling system acceptance with 
an acceptance of pictures row by row using an automaton model over ordinal words of length 
$\om^2$. 
We give in this paper a solution to this problem, showing 
that all languages of infinite pictures which are accepted row by row 
by B\"uchi or Choueka 
automata reading words of 
length $\om^2$ are B\"uchi recognized by a finite tiling system, but 
the converse is not true. 
We give also the  answer to two other questions which were raised by Altenbernd, Thomas and 
W\"ohrle, 
showing that it is undecidable whether a B\"uchi 
recognizable language of infinite pictures is E-recognizable (respectively, A-recognizable). 


\end{abstract}
\textlineskip
\section{Introduction}

 In a recent paper \cite{atw}, Altenbernd, Thomas and 
W\"ohrle have considered acceptance of 
languages of infinite two-dimensional words (infinite pictures) by finite tiling systems,  
with the usual acceptance conditions, such as  the 
B\"uchi and Muller ones,  firstly  used for acceptance of infinite words.
This way they extended the classical theory of recognizable languages of finite pictures, 
  \cite{gr},  to the case of infinite pictures.  
\nl  On the other hand 
automata reading  ordinal words have been first considered by B\"uchi in order to 
study the decidability of the monadic second order theory on  
 countable ordinals. 
In particular he defined automata reading words of length $\om^2$, \cite{bu62,bs}. 
Another model 
of automaton reading words of length $\om^2$ has been studied by Choueka in \cite{cho} 
and  it has been 
shown by Bedon that these two models are equivalent \cite{bed,bedb}. 
They accept the so called regular 
$\om^2$-languages which can also be defined by generalized regular expressions, see also the 
work of Wojciechowski \cite{w2,w3}. 
\nl  In \cite{atw} the authors asked for  comparing the tiling system acceptance with 
an acceptance of pictures row by row using an automaton model over ordinal words of length 
$\om^2$. 
\nl   We give in this paper a solution to this problem, showing 
that the class  of 
 languages of infinite pictures which are accepted by B\"uchi  
automata reading words of 
length $\om^2$ is strictly included in the class  of 
 languages of infinite pictures which are B\"uchi-recognized by some finite tiling system. 
\nl  
Another way to compare these two classes is to compare the topological complexity 
of languages in each of them, with regard to the Borel and projective hierarchies. 
We then determine the topological complexity of B\"uchi-recognized 
languages of infinite pictures. This way we show 
that B\"uchi tiling systems have a much greater accepting power than 
automata over ordinal words of length $\om^2$. 
\nl Using topological arguments,  we 
give also  the  answer to two questions raised in  \cite{atw}, 
showing that it is undecidable whether a B\"uchi 
recognizable language of infinite pictures is E-recognizable (respectively, A-recognizable). 
 For that purpose we  use  a very similar technique as in a recent paper where  
we have proved several undecidability results for 
infinitary rational relations \cite{rel-dec}. 
\nl  The paper is organized as follows. In section 2 we recall  basic definitions for 
pictures and tiling systems.  B\"uchi automata reading words of length $\om$ or $\om^2$ 
are introduced in section 3.  We compare the two modes of acceptance  in section 4.  
 Undecidability results are proved in section 5.

\section{Tiling Systems}

 Let $\Si$ be a finite alphabet and $\#$ be a letter not in $\Si$ and let 
$\hat{\Si}=\Si \cup \{\#\}$. If $m$ and $n$ are two integers $>0$ or if $m=n=0$,  
a   picture of size $(m, n)$ over $\Si$ 
is a function $p$ from $\{0, 1, \ldots , m+1\} \times \{0, 1, \ldots , n+1\}$ 
into $\hat{\Si}$ such that 
$p(0, i)=p(m+1, i)=\#$ for all integers $i\in \{0, 1, \ldots , n+1\}$ and 
$p(i, 0)=p(i, n+1)=\#$ for all integers $i\in \{0, 1, \ldots , m+1\}$ and 
$p(i, j) \in \Si$ if $i \notin \{0, m+1\}$ and $j \notin \{0, n+1\}$. The empty picture 
is the only picture of size $(0, 0)$ and is denoted by $\lambda$. Pictures of  size 
$(n, 0)$ or $(0, n)$, for $n>0$, are not defined. $\Si^{\star, \star}$ is the set of 
pictures over $\Si$. A picture language $L$ is a subset of $\Si^{\star, \star}$. 

\hs An $\om$-picture over $\Si$ 
is a function $p$ from $\om \times \om$ into $\hat{\Si}$ such that $p(i, 0)=p(0, i)=\#$ 
for all $i\geq 0$ and $p(i, j) \in \Si$ for $i, j >0$. $\Si^{\om, \om}$ is the set of 
$\om$-pictures over $\Si$. An $\om$-picture language $L$ is a subset of $\Si^{\om, \om}$. 
\nl   For $\Si$ a finite alphabet we call  $\Si^{\om^2}$  the set of functions 
from $\om \times \om$ into $\Si$. So the set $\Si^{\om, \om}$ of $\om$-pictures over 
$\Si$ is a strict subset of $\hat{\Si}^{\om^2}$.  

\hs We shall say that, for each integer $j\geq 1$,  the $j^{th}$ row of an $\om$-picture 
$p\in \Si^{\om, \om}$ is the infinite word $p(1, j).p(2, j).p(3, j) \ldots$ over $\Si$ 
and the $j^{th}$ column of $p$ is the infinite word $p(j, 1).p(j, 2).p(j, 3) \ldots$ 
over $\Si$. 
\nl As usual,  one can imagine that,  for integers $j > k \geq 1$, the $j^{th}$ column of $p$ 
is on the right of the $k^{th}$ column of $p$ and that the $j^{th}$ row of $p$ is 
``above" the $k^{th}$ row of $p$. This representation will be used in the sequel. 

\hs We introduce now tiling systems as in the paper \cite{atw}. 
\nl A tiling system is a tuple $\mathcal{A}$=$(Q, \Si, \Delta)$, where $Q$ is a finite set 
of states, $\Si$ is a finite alphabet, $\Delta \subseteq (\hat{\Si} \times Q)^4$ is a finite set 
of tiles. 
\nl A B\"uchi tiling system is a pair $(\mathcal{A},$$ F)$ 
 where $\mathcal{A}$=$(Q, \Si, \Delta)$ 
is a tiling system and $F\subseteq Q$ is the set of accepting states. 
\nl A Muller tiling system is a pair $(\mathcal{A}, \mathcal{F})$ 
where $\mathcal{A}$=$(Q, \Si, \Delta)$ 
is a tiling system and $\mathcal{F}$$\subseteq 2^Q$ is the set of accepting sets of states.

\hs  Tiles are denoted by 
$
\left ( \begin{array}{cc}  (a_3, q_3) & (a_4, q_4)
\\ (a_1, q_1) & (a_2, q_2) \end{array} \right ) \mbox{ with } a_i \in \hat{\Si} \mbox{ and } 
q_i \in Q, 
$

\hs and in general, over an alphabet $\Ga$, by 
$
\left ( \begin{array}{cc} b_3 & b_4 
\\ b_1 & b_2 \end{array} \right ) \mbox{  ~~~~~~with  }  b_i \in \Ga . 
$

\noi We will indicate a combination of tiles by: 
\begin{displaymath}
\left ( \begin{array}{cc}  b_3 & b_4
\\ b_1 & b_2 \end{array} \right ) \circ  
\left ( \begin{array}{cc} b'_3 & b'_4
\\ b'_1 & b'_2  \end{array} \right ) = 
\left ( \begin{array}{cc} (b_3, b'_3) & (b_4, b'_4) 
\\ (b_1, b'_1) & (b_2, b'_2) \end{array} \right ) 
\end{displaymath}

\noi A run of a tiling system $\mathcal{A}$=$(Q, \Si, \Delta)$ over a (finite) 
picture $p$ of size $(m, n)$ over $\Si$ 
is a mapping $\rho$  from $\{0, 1, \ldots , m+1\} \times \{0, 1, \ldots , n+1\}$ 
into $Q$ such that for all $(i, j) \in \{0, 1, \ldots , m\} \times \{0, 1, \ldots , n\}$ 
with $p(i, j)=a_{i, j}$ and $\rho(i, j)=q_{i, j}$ we have 
\begin{displaymath}
\left ( \begin{array}{cc}  a_{i, j+1} & a_{i+1, j+1} 
\\  a_{i, j}  & a_{i+1, j}   \end{array} \right ) \circ  
\left ( \begin{array}{cc} q_{i, j+1} & q_{i+1, j+1} 
\\ q_{i, j}  & q_{i+1, j}   \end{array} \right ) \in \Delta . 
\end{displaymath}

\noi A run of a tiling system $\mathcal{A}$=$(Q, \Si, \Delta)$ over an 
$\om$-picture $p \in \Si^{\om, \om}$ 
is a mapping $\rho$  from $\om \times \om$ 
into $Q$ such that for all $(i, j) \in \om \times \om$ 
with $p(i, j)=a_{i, j}$ and $\rho(i, j)=q_{i, j}$ we have 
\begin{displaymath}
\left ( \begin{array}{cc}  a_{i, j+1} & a_{i+1, j+1} 
\\  a_{i, j}  & a_{i+1, j}   \end{array} \right ) \circ  
\left ( \begin{array}{cc} q_{i, j+1} & q_{i+1, j+1} 
\\ q_{i, j}  & q_{i+1, j}   \end{array} \right ) \in \Delta . 
\end{displaymath}

\noi We  now recall  acceptance of finite or  infinite pictures by tiling systems:

\begin{Deff} Let $\mathcal{A}$=$(Q, \Si, \Delta)$ 
be a tiling system, $F\subseteq Q$ and $\mathcal{F}$$\subseteq 2^Q$. 
\begin{itemize}

\ite
 The picture language recognized by $\mathcal{A}$ 
is the set of pictures $p \in \Si^{\star, \star}$ such that there is some run $\rho$ of 
$\mathcal{A}$ on $p$. 

\ite The $\om$-picture language A-recognized  (respectively, E-recognized, B\"uchi-recognized) 
by 
$(\mathcal{A},$$ F)$ 
is the set of $\om$-pictures $p \in \Si^{\om, \om}$ such that there is some run $\rho$ of 
$\mathcal{A}$ on $p$ and $\rho(v) \in F$ for all  
(respectively, for at least one,  for infinitely many) $v\in \om^2$. 

\ite The $\om$-picture language Muller-recognized  by $(\mathcal{A}, \mathcal{F})$ is 
the set of $\om$-pictures $p \in \Si^{\om, \om}$ such that there is some run $\rho$ of 
$\mathcal{A}$ on $p$ and $Inf(\rho) \in \mathcal{F}$ where $Inf(\rho)$ is the set of states 
occurring infinitely often in $\rho$. 
\end{itemize}
\end{Deff}

\noi As stated in \cite{atw}, an $\om$-picture language $L \subseteq \Si^{\om, \om}$ is 
recognized by a B\"uchi tiling system if and only if it 
is recognized by a  Muller tiling system.  
\nl We shall denote $TS(\Si^{\om, \om})$ the class of languages $L \subseteq \Si^{\om, \om}$ 
which are recognized by some  B\"uchi (or Muller) tiling system. 
 
\section{B\"uchi Automata}  

 We shall assume  the reader to be familiar with the elementary theory 
of countable ordinals, which may be found in \cite{sier}. In fact we shall only 
need in this section to consider ordinals smaller than $\om^2+1$.
\nl Let $\Si$ be a finite  alphabet, and  $\alpha$ be a countable  ordinal. 
An $\alpha$-word $x$
 (word of length $\alpha$) over the alphabet $\Si$ is an  $\alpha$-sequence
(sequence of length $\alpha$) of letters in $\Si$. It will be denoted by 
$(x(i))_{0\leq i<\alpha}=x(0).x(1).x(2)\ldots x(i) \ldots$ ,  
where for all $i$, $0\leq i<\alpha$, $x(i)$ is a letter in $\Si$. 
\nl For   an ordinal $\alpha \geq \omega$,  
 the set of  $\alpha$-words over 
 $\Si$ will be denoted by $\Si^\alpha$. An  $\alpha$-language over $\Si$ is a subset of 
 $\Si^\alpha$. 

\hs We assume now that the reader has some familiarity with the notion of B\"uchi and Muller 
automata reading infinite words, \cite{tho,sta,pp}.

\begin{Deff}
 A B\"uchi automaton  is a 5-tuple
 $\mathcal{A}$= $(\Sigma, Q, q_0 ,\Delta , F )$ 
where  $Q$ is a finite set of states, $q_0 \in Q$ is the 
 initial  state, $\Delta \subseteq Q\times\Sigma\times Q $ is the  transition relation, 
and $F \subseteq Q$ is the set of final states.
\nl A run of $\mathcal{A}$ on the  $\omega$-word  $\sigma \in \Sio$ 
 is an  $\omega$-sequence  $x \in Q^\om$  such 
that $x(0)=q_0$ and  $(x(i),\sigma(i),x(i+1))\in\Delta$ for  $i \geq 0$. 
\nl The run is called successful if  $ Inf ( x ) \cap F\neq \emptyset $, where 
$Inf( x ) $
 is the set of elements of  $Q$ which appear infinitely often in the $\omega$-sequence 
 $x$.
\nl An  $\omega$-word $\sigma \in \Sio$ is accepted by $\mathcal{A}$ 
if there exists a successful run of  $\mathcal{A}$ 
on   $\sigma$. 
\nl  $L_\omega (\mathcal{A})$=$\{\sigma \in \Sigma^\omega \mid \mathcal{A} \mbox{ accepts }
\sigma \} $ is the $\omega$-language recognized by   $\mathcal{A}$.
\end{Deff}

\noi A Muller automaton is defined in a similar way except that $F$ is replaced by a set 
$\mathcal{F}$ $\subseteq 2^Q$ of accepting sets of states and that a run $x \in Q^\om$ 
on an $\omega$-word  $\sigma \in \Sio$ is said to be successful iff 
 $Inf( x ) \in \mathcal{F}$. 
\nl B\"uchi and Muller automata accept the same 
class of \ol s: the class of regular \ol s which is the $\om$-Kleene closure of the class 
of regular finitary languages.  
It follows from Mac Naughton's Theorem that each regular $\om$-language is also 
accepted by a deterministic Muller automaton, \cite{tho,sta,pp}. 

\hs In order to define an automaton  reading ordinal  words of length
$\geq\omega$, we must add to the  automaton  a  transition relation 
for limit steps: after the reading of a word which length is a limit  ordinal,  
the state of the automaton will depend on the set of states 
which cofinally appeared during the run of the automaton,  
\cite{bs,hem,bed}.  We shall give the following definition in the general 
case of  automata 
reading  ordinal words  but in fact we shall only need in the sequel 
the notion of automata reading words of length $\om$ or $\om^2$. 

\begin{Deff} An ordinal B\"uchi automaton is a  sextuple
 $\mathcal{A}$=$(\Sigma, Q, q_0, \Delta, \gamma, F)$ where:
$\Si $ is a finite alphabet,
$Q$ is a finite set of states,
$q_0 \in Q$ is the   initial state,
$\Delta \subset Q \times \Sigma \times Q$ is the transition relation, and 
$\gamma \subset P(Q) \times Q$ is the transition relation for limit steps.
\end{Deff}

\noi $\Sigma, Q, q_0, \Delta$ and $F$ keep the same  meaning as before,
 the meaning of $\gamma$ is given by the following definition:

\begin{Deff} A run of the ordinal  B\"uchi automaton  
 $\mathcal{A}$=$(\Si, Q, q_0, \Delta , \gamma, F)$,  reading the word  $\sigma \in \Si^\alpha$,  
 is an ($\alpha +1$)-sequence of states  $x$  defined by:
$x(0)=q_0$ and for  $i<\alpha$, $(x(i), \sigma(i), x(i+1))\in \Delta$ and for 
a limit ordinal  $i$: $(Inf(x,i),x(i)) \in \gamma$, where 
$$Inf(x, i)=\{ q\in Q \mid \fa \mu <i, \exists \nu<i  \mbox{ such that }
 \mu<\nu  \mbox{  and  } x(\nu)=q \} $$
\noi is the set of states which cofinally appear during the reading 
of the  $i$ first letters of  $\sigma$. 
\nl A run  $x$  of the automaton  $\mathcal{A}$ over the word 
 $\sigma$
is called successful if $x(\alpha) \in F$. A word $\sigma \in \Si^\alpha$ 
is accepted  by  $\mathcal{A}$ if there exists a  successful run of 
$\mathcal{A}$ over $\sigma$. We denote 
 $L_\alpha(\mathcal{A})$ the set of words of length  $\alpha$ which are accepted by 
 $\mathcal{A}$.

\end{Deff}

\noi In particular the above definition  
 provides a notion of automata reading words of length $\om^2$. 
Later Choueka defined another class of automata reading words of length $\om^2$ 
(and even $\om^n$ for an integer $n\geq 2$) now called Choueka automata \cite{cho}.  
Bedon proved  that these two classes of automata accept the same class of 
$\om^2$-languages, the class of regular $\om^2$-languages which can be also defined by 
$\om^2$-regular expressions \cite{bed,bedb}.

\begin{Rem} When we consider only finite words, the language accepted by an 
ordinal  B\"uchi automaton 
is a  rational language. And an  $\omega$-language
is accepted by an ordinal  B\"uchi automaton if and only if it is accepted
by a Muller automaton hence also by a  B\"uchi automaton.
\end{Rem}

\noi We shall use in the sequel another way of generating regular $\om^2$-languages 
which is given by the following proposition. We shall reprove this result although it already 
appeared in \cite{hem} and has been  also proved in \cite{loc}.

\begin{Pro}\label{sub}
An $\om^2$-language $L \subseteq \Si^{\om^2}$  is  regular iff it is obtained from a regular 
\ol~ $R \subseteq \Ga^\om$  
by substituting in every $\om$-word $\sigma \in R$ 
a regular \ol~ $L_a  \subseteq \Si^\om$ to each letter $a\in \Ga$. 
\end{Pro}

\proof{ 
Let $\mathcal{A}$=$(\Sigma, Q, q_0, \Delta, \gamma, F) $ be an ordinal  B\"uchi automaton, 
and let  $L_{\omega^2}(\mathcal{A})$ be   the $\omega^2$-language recognized by  $\mathcal{A}$.
\nl Consider the reading of a word  $\sigma \in \Si^{\om^2}$ by $\mathcal{A}$: 
After the reading of the first $\omega$ letters, 
 $\mathcal{A}$ is in state  $x(\omega)$, after the reading of  $\omega.2$ 
 letters, $\mathcal{A}$ is in state  $x(\omega.2)$ and so on.  
\nl For $q_i\in Q, q_j\in Q $ and $E\subseteq Q$, we denote by $L(q_i,q_j,E)$ 
 the $\omega$-language of words  $u \in \Sio$ such that there exists a reading of $u$ by 
$\mathcal{A}$,  beginning in state  $q_i$,  ending  
 in state $q_j$ after the reading of $u$,  and  going through the set of 
  states $E$ (including $q_i $  and $q_j$). 
\nl We easily see that these \ol s are recognized by Muller automata therefore
 also by B\"uchi automata.

\hs Consider now the  new alphabet:
$$\Gamma=Q\times Q\times P(Q)=\{ (q_i,q_j, E) \mid q_i\in Q, q_j\in Q, E\subseteq Q\}$$
\noi and let $R \subseteq \Ga^\om$ containing an $\om$-word $\sigma \in \Ga^\om$ if and only if 
$\sigma$ satisfies the two following properties: 
\nl $(1)$. The first letter of $\sigma$ is in the form
$(q_0,q , E)$ and each letter $(q_i,q_j,E)$ is followed by a letter 
 $(q_j,q,G)$ with  $q\in Q, G \subseteq Q$. 
\nl $(2)$.  The set 
 $$X=\{ q\in Q \mid \mbox{  some letter } (q_i,q_j,G)  \mbox{  appears   
infinitely often in } \sigma \mbox{ and  } q\in G \}$$
\noi  satisfies  
$(X, q_f)\in\gamma$ for some  $q_f\in F$.

\hs  $R$ is a regular   \ol~ and  if we substitute  in $R$ 
the $\omega$-language  $L(q_i,q_j,E)$ to  each letter $(q_i,q_j, E)$, we obtain  
the $\omega^2$-language recognized by   $\mathcal{A}$, i.e. $L_{\omega^2}(\mathcal{A})$.

\hs We have then proved one implication of Proposition \ref{sub}. In fact we shall only 
need in the sequel this implication. 
\nl We just mention that the converse can be easily proved by using regular expressions 
defining regular \ol s and regular  $\om^2$-languages.  }

\hs We have now to define precisely the acceptance of infinite pictures  row by row by 
an automaton model over ordinal words of length $\om^2$.  
\nl To an infinite picture $p  \in  \Si^{\om, \om}$ we associate an $\om^2$-word 
$\bar{p} \in \Si^{\om^2}$ which is defined by $\bar{p}(\om.n + m)=p(m+1, n+1)$ 
for all integers $n, m \geq 0$. 
\nl This can be extended to languages of infinite pictures: for $L \subseteq \Si^{\om, \om}$ 
we denote $\bar{L}=\{\bar{p} \mid p \in  L \}$ so $\bar{L}$ 
 is an $\om^2$-language over $\Si$. 
 
\hs We can now set the following definition: 

\begin{Deff}
A language  of infinite pictures $L \subseteq \Si^{\om, \om}$ 
is accepted row by row by an ordinal B\"uchi 
automaton  if and only if 
the  $\om^2$-language $\bar{L}$ is regular. 
\nl We shall denote $BA(\Si^{\om, \om})$ the class of languages $L \subseteq \Si^{\om, \om}$ 
such that $\bar{L}$ is regular, i.e. is  accepted by an ordinal  B\"uchi automaton. 
\end{Deff}

\begin{Rem}
We have defined the $\om^2$-word 
$\bar{p}$ without the letters $\#$ appearing in the infinite picture $p$. It is easy to see 
that this does not change the notion of acceptance of a language of  infinite pictures 
  row by row by an ordinal  B\"uchi automaton. 
\end{Rem}

\section{Comparison of The Two Modes of Acceptance} 

 We can now state our main result. 

\begin{The}\label{mainthe}
Every language of infinite pictures which is accepted row by row by an ordinal  B\"uchi  
automaton is B\"uchi-recognized by some finite tiling system, but 
the converse is not true. 

\end{The}

\noi We are going to split the proof of Theorem \ref{mainthe} into the two following lemmas. 

\begin{Lem}
Every language of infinite pictures which is accepted row by row by an ordinal  B\"uchi  
automaton is  B\"uchi-recognized by some finite tiling system. 
\end{Lem}

\proof{  Let  $L \subseteq \Si^{\om, \om}$ be a language of infinite pictures which 
is accepted row by row by an ordinal  B\"uchi  automaton,  i.e. such that 
the  $\om^2$-language $\bar{L}$ is regular. 

\hs By Proposition \ref{sub}, the $\om^2$-language   $\bar{L}$  is obtained from a regular 
\ol~ $R \subseteq \Ga^\om$, where $\Ga=\{a_1, a_2, \ldots , a_n\}$ is a finite alphabet,   
by substituting in every $\om$-word $\sigma \in R$ 
a regular \ol~ $R_{i} \subseteq \Si^\om$ to each letter $a_i \in \Ga$. 

\hs Let $\mathcal{A}$= $(\Ga, Q, q_0 ,\Delta , F )$ be a B\"uchi automaton accepting the 
 regular $\om$-language $R$ and,  
for each integer $i \in [1; n]$, let 
$\mathcal{A}$$^{i}$= $(\Sigma, Q^{i}, q_0^{i} ,\Delta^{i} , F^{i} )$ 
 be a B\"uchi automaton accepting the regular $\om$-language $R_{i}$. 
 We  assume, without loss of generality,  that for all integers $i, j \in [1; n]$,  
$Q^{i} \cap Q^{j} = \emptyset$ and $Q^{i} \cap Q = \emptyset$.

\hs We shall describe the behaviour of a tiling system 
$\mathcal{T}$=$(K, \Si, \Delta^{\mathcal{T}})$ which will accept infinite pictures 
$p \in L$ with a Muller acceptance condition. 
\nl A run $\rho$ of $\mathcal{T}$ on an $\om$-picture $p \in L$ will guess, for each integer 
$j\geq 1$, an integer $i_j \in \{1, 2, \ldots, n\}$ such that  the 
$j^{th}$ row $p_j$ of $p$ is in $R_{i_j}$. It will then check that for all $j\geq 1$ 
the $\om$-word $p_j$ is in  $R_{i_j}$ and that 
the $\om$-word $a_{i_1}.a_{i_2} \ldots a_{i_j} \ldots $ is in $R$. 

\hs We are going now to describe informally 
a run $\rho$ of $\mathcal{T}$ over an infinite picture 
$p\in \Si^{\om, \om}$. 
 
\hs Each state of $\mathcal{T}$, i.e. each element of $K$, will 
consist of five components.

\hs The first component of a state of $\mathcal{T}$ 
is an integer $i_j \in \{1, 2, \ldots , n\}$. 
\nl It will be used 
to guess that the $\om$-word $p_j=p(1, j).p(2, j).p(3, j) \ldots$, forming the $j^{th}$ row 
of the picture $p$, is in the regular $\om$-language $R_{i_j}$. 
\nl This  first component will be constant on every row of the  run $\rho$  and 
will be propagated horizontally.

\hs The second component is an element of $\cup_{1\leq i\leq n} Q^{i}$. 
\nl If on the  $j^{th}$ row  the first component of the state 
is  equal to  $i_j$ then the second component 
on this row will be in  $Q^{i_j}$.  
It is used to simulate (by horizontal propagation) a run $\alpha_j$ 
of the B\"uchi automaton $\mathcal{A}$$^{i_j}$ on the $\om$-word 
$p_j$ forming the $j^{th}$ row 
of $p$. 
\nl So the projection of $\rho_j=\rho(1, j).\rho(2, j)\ldots$ on the second component 
of  states  will be  equal to $\alpha_j$. 

\hs In order to check that, for all integers $j\geq 1$, the $\om$-word $p_j$ is in the 
regular $\om$-language $R_{i_j}$, $\mathcal{T}$ has  to check that each run $\alpha_j$ is 
successful, i.e. that $Inf(\alpha_j)\cap F^{i_j} \neq \emptyset$, or equivalently that  
some state of $F^{i_j}$ appears 
infinitely often in the second component (of the state)  on the $j^{th}$ row. 

\hs This can be done in the following way. One can imagine an ant which moves  on the picture 
$p$, but only horizontally from the left to the right or vertically.  The  movement of the ant 
will be indicated by the third component of the state which will be an element 
of $\{B, a, a_d\}$. 
\nl Letters $a$, $a_d$ will represent the trajectory of the ant and the blank symbol $B$ will 
be  used elsewhere. The letter $a_d$ will be only used when the ant goes down vertically 
on the picture. 
\nl We shall need also the fourth component of the state of $\mathcal{T}$ 
which will be an element 
of $\{B, \star, \star_1\}$. 

\hs The walk of the ant begins at the intersection of the first row and the first column 
of $p$, i.e. at the place of the letter $p(1, 1)$ of $p$. 
\nl At the beginning of this walk, the ant  moves horizontally to the right on the first row 
(this way is marked by  an $a$ on the third component of the state)  until it meets an 
element  $q_1 \in F^{i_1}$ on the second component of the state. 
\nl There is also  a mark $\star$ on  the first row 
which is propagated to the right following the movement  of the ant.   

\hs If the ant meets an element  $q_1 \in F^{i_1}$ on the second component of the state, 
then the  mark $\star$ is transferred on the 
second row just above it (on the same column) but with an indice $1$, so it becomes 
$\star_1$. 
\nl   This mark $\star_1$ 
will be next forwarded horizontally to the right but without the indice $1$.

\hs The ant then goes down vertically until it reaches the first row. In that special 
beginning of its walk, it is already on the first row! 
\nl  Next the ant moves again to the right on the first row, until it meets an 
element  $q_2 \in F^{i_1}$ on the second component of the state. At that point 
it goes up on the second row (which is marked by $\star$ on the fourth component) 
and moves to the right on this row until it meets an  
element  $q_3 \in F^{i_2}$ on the second component of the state. 
\nl At that point  the  mark $\star$ is transferred on the 
third row just above it (on the same column) but with an indice $1$, so it becomes 
$\star_1$. This mark $\star_1$ 
will be next forwarded horizontally to the right but without the indice $1$. 
\nl The mark $\star$ is now on the third row and it indicates that the ant will have 
to check successively the three first rows at next 
ascending moves. 

\hs The ant then goes down vertically until it reaches the first row. These movements 
will be indicated by the letter $a_d$ on the third component of the state. 
Once on the first row its trajectory is again marked by the letter $a$. 
It moves to the right, looking for some state of $F^{i_1}$ on the first row, next goes up, 
moves to the right, looking for some state of $F^{i_2}$ on the second row,  
next goes up, moves to the right, again looking for some state of $F^{i_3}$ on the third row. 
\nl  This way it checks successively  the first row, then the second row, and the third 
row (marked with $\star$), looking each time for an element of $F^{i_j}$ on the  $j^{th}$  row. 
When it meets an 
element  $q_6 \in F^{i_3}$ on the second component on the third row,  
it transfers the  mark $\star$ (with an indice, so it becomes 
$\star_1$) just above it. This mark $\star$ 
will be next forwarded horizontally to the right, without the indice $1$. 
\nl The mark $\star$ is now on the fourth  row and it indicates that the ant will have 
to check successively the four  first rows at next 
ascending moves. 

\hs  The ant then goes down vertically until it reaches the first row and so on \ldots

\hs We can see that if  the mark $\star_1$ appears infinitely often, it appears one 
time on each row, and this means that the  ant has successively checked the first row, 
then the two first rows, then the three first rows, \ldots , then the $n$ first rows, \ldots, 
looking each time for an element of $F^{i_j}$ on the  $j^{th}$  row. 
\nl This implies  that, for a  given  $j^{th}$  row, the  ant  has checked that some 
element of $Q^{i_j}$ appears infinitely often on the second component of the state, hence the 
 $\om$-word $p_j=p(1, j).p(2, j).p(3, j) \ldots$  is in the regular $\om$-language $R_{i_j}$. 
\nl Conversely if for all integers $j \geq 1$ the $\om$-word $p_j$ is in $R_{i_j}$, then 
there are some successful runs $\alpha_j$ of the  
B\"uchi automata $\mathcal{A}$$^{i_j}$ on the $\om$-words $p_j$ such that 
the above defined movements 
of the ant make the mark $\star_1$ to appear infinitely often. 
\nl  Notice that the blank symbol $B$ appears on the fourth component of the state whenever 
neither $\star$ nor $\star_1$ is used as explained above. 

\hs $\mathcal{T}$ has now to check that the integers $i_j$, $j \geq 1$, are such that  
the $\om$-word $a_{i_1}.a_{i_2} \ldots a_{i_j} \ldots $ is in $R$. 
The fifth component of states of  $K$ is used for that purpose. 
On the first column this fifth component is an element of $Q$ and 
is used to simulate, by vertical 
propagation,  a run $\alpha$ of 
$\mathcal{A}$ on  the $\om$-word $a_{i_1}.a_{i_2} \ldots a_{i_j} \ldots$ . 
\nl This means that the projection of $\rho(1, 1).\rho(1, 2).\rho(1, 3)\ldots$ 
on this fifth component  will be  equal to $\alpha$. 
\nl On the other columns the fifth component will be simply the blank symbol $B$. 

\hs We have seen that the set of states of the tiling system $\mathcal{T}$ will be: 
$$K = \{1, 2, \ldots , n\} \times \cup_{1\leq i\leq n} Q^{i} \times \{B, a, a_d\}  
\times \{B, \star, \star_1\} \times (\{B\}\cup Q)$$

\noi and one can 
define a set of tiles $\Delta^{\mathcal{T}}$ such that corresponding runs 
of the tiling system 
$\mathcal{T}$=$(K, \Si, \Delta^{\mathcal{T}})$  are described  informally as above.  

\hs A run $\rho$ will be successful if and only if the mark $\star_1$ 
appears infinitely often on the fourth component of $\rho(v)$ and some state 
$q\in F$ appears infinitely often on the fifth component of $\rho(v)$,  for 
$v\in \om^2$. 
\nl This acceptance condition may be written as a Muller condition. 
As stated in \cite{atw} any language of $\om$-pictures which is Muller recognizable by 
a  tiling system is also B\"uchi recognizable by 
a  tiling system.  
 } 

\begin{Lem}
There exists a  B\"uchi-recognizable language of infinite pictures 
which is  not accepted row by row by any ordinal  B\"uchi  automaton. 
\end{Lem}

\proof{ 
The class of languages of 
infinite pictures which are B\"uchi-recognizable by tiling systems  
is not closed under complement, \cite{atw}. There 
exists a language $T \subseteq \Si^{\om, \om}$ 
of infinite pictures, (where $\Si$ is a finite alphabet),  
which is B\"uchi-recognizable by a tiling system 
but such that its complement is not B\"uchi-recognizable by any tiling system. 
\nl Then  $\bar{T}$ cannot be a regular $\om^2$-language. Indeed  otherwise 
its complement would be also a regular $\om^2$-language because the class of 
regular $\om^2$-languages is closed under complement, \cite{bed,bedb}. 
The preceding  proof would imply that the complement of $T$ would be  also 
B\"uchi-recognizable by a tiling system, towards a contradiction. 
 } 

\hs 
Theorem \ref{mainthe}  expresses  that the class $BA(\Si^{\om, \om})$ is strictly included 
in the class $TS(\Si^{\om, \om})$.  We shall see in the next section that one cannot decide 
whether a language $L \in TS(\Si^{\om, \om})$ is in $BA(\Si^{\om, \om})$. 

\hs We are going now to  compare  the topological complexity 
of languages  in the classes $TS(\Si^{\om, \om})$ and $BA(\Si^{\om, \om})$. 
\nl  From now on we shall assume  
that the reader is familiar with basic notions of topology and with 
the Borel and projective hierarchies 
on a space $\Si^{\om}$, (where $\Si$ is a  finite alphabet  having at least two letters),  
equipped with the Cantor topology, 
see for example \cite{lt,sta,pp,kec}. 

\noi  We recall that  a subset of $\Si^\om$ is a Borel set of rank $\alpha$, for 
a countable ordinal $\alpha$,  iff 
it is in ${\bf \Si^0_{\alpha}}\cup {\bf \Pi^0_{\alpha}}$ but not in 
$\bigcup_{\gamma <\alpha}({\bf \Si^0_\gamma }\cup {\bf \Pi^0_\gamma})$. 
\nl Recall also the notion of completeness with regard to reduction by continuous functions. 
For a countable ordinal  $\alpha \geq 1$, a set $F\subseteq \Si^\om$ is said to be 
a ${\bf \Si^0_\alpha}$  (respectively,  ${\bf \Pi^0_\alpha}$, ${\bf \Si^1_1}$)-complete set 
iff for any set $E\subseteq \Ga^\om$  (with $\Ga$ a finite alphabet): 
 $E\in {\bf \Si^0_\alpha}$ (respectively,  $E\in {\bf \Pi^0_\alpha}$,  $E\in {\bf \Si^1_1}$) 
iff there exists a continuous function $f: \Ga^\om \ra \Si^\om$ such that $E = f^{-1}(F)$.

\hs  For $\Ga$ a finite alphabet having at least two letters, the 
set $\Ga^{\om \times \om}$ of functions  from $\om \times \om$ into $\Ga$ 
is usually equipped with the product topology  of the discrete 
topology on $\Ga$. 
This topology may be defined 
by the following distance $d$. Let $x$ and $y$  in $\Ga^{\om  \times \om}$ 
such that $x\neq y$, then  
$$ d(x, y)=\frac{1}{2^n} ~~~~~~~\mbox{   where  }$$
$$n=min\{p\geq 0 \mid  \exists (i, j) ~~ x(i, j)\neq y(i, j) \mbox{ and } i+j=p\}.$$

\noi  Then the topological space $\Ga^{\om \times \om}$ is homeomorphic to the 
topological space $\Ga^{\om}$, equipped with the Cantor topology.  
Borel subsets  of   $\Ga^{\om  \times \om}$ are defined from open 
subsets  as in the case of the topological space $\Ga^\om$. 
Analytic  subsets  of   $\Ga^{\om  \times \om}$ are obtained as projections on 
$\Ga^{\om \times \om}$
 of Borel subsets of the product space  $\Ga^{\om \times \om} \times \Ga^\om$.
\nl  The set $\Si^{\om, \om}$ of $\om$-pictures over $\Si$, 
viewed as a topological subspace of $\hat{\Si}^{\om \times \om}$, 
 is easily seen to be homeomorphic to the topological space $\Si^{\om \times \om}$, 
via the mapping 
$\varphi: \Si^{\om, \om} \ra \Si^{\om \times \om}$ 
defined by $\varphi(p)(i, j)=p(i+1, j+1)$ for all 
$p\in \Si^{\om, \om}$ and $i, j \in \om$.

\hs The topological complexity of 
languages of infinite pictures,   
 accepted row by row  by ordinal B\"uchi automata, 
is given by the following result  
which is stated in \cite{dfr}.  

\begin{Pro}[\cite{dfr}]
Let $L \subseteq \Si^{\om, \om}$ be a language of infinite pictures which 
is accepted row by row by an ordinal B\"uchi 
 automaton. Then $L$ is a Borel set of rank 
 smaller than or equal to 5. 
\end{Pro}

\noi This result can be easily proved, using Proposition \ref{sub} and the fact that every
regular $\om$-language $R \subseteq \Ga^\om$ 
is a boolean combination 
of arithmetical  $\Pi_2$-sets, hence a $\Delta_3$-set, so  is 
definable in first order arithmetic 
by  some first order $\Si_3$-sentence and also by some first order $\Pi_3$-sentence.
 One can then show that every regular $\om^2$-language is 
defined in first order arithmetic by some first order $\Si_5$-sentence hence is a Borel 
set of rank smaller than or equal to 5.

\hs On the other side it has been proved in \cite{atw} that there exist some 
${\bf \Si^1_1}$-complete, hence non Borel, B\"uchi recognizable language of $\om$-pictures.  
 The two following lemmas will provide an alternative proof of this result and will 
be also useful  to determine the Borel ranks of languages  in $TS(\Si^{\om, \om})$. 

\hs   For an \ol~  $L \subseteq \Sio$  we  denote  $L^B$  the language of infinite pictures 
$p \in \Si^{\om, \om}$ such that  the first row of $p$ is in $L$ and the other rows are 
labelled with the letter $B$ which is assumed to belong to $\Si$.

\begin{Lem}\label{lem2'}
If   $L \subseteq \Sio$ is 
accepted by some Turing machine with a B\"uchi acceptance 
condition, then $L^B$ is B\"uchi recognizable by a finite tiling system. 
\end{Lem}

\proof{  
Let $L \subseteq \Sio$ be an \ol~ accepted by some Turing machine $T$ 
with a B\"uchi acceptance condition. 
\nl  We assume that the Turing machine has a single semi-infinite tape, 
with one reading head which may 
also write on the tape. $Q$ is the set of states of $T$, $q_0$ is the initial state and 
$F \subseteq Q$ is the set of accepting states.  The input alphabet of $T$ is $\Si$ and its 
working alphabet is $\Ga \supseteq \Si$. 
\nl It has been proved by Cohen and Gold that one can consider only such restricted model 
of Turing machine \cite{cg}. 
\nl An instantaneous configuration of $T$ is given by an infinite word 
$u.q.v$ where $u\in \Gas$, $q\in Q$,  $v\in \Gao$, and the first letter of $v$ is the one 
scanned by the head of $T$. 
\nl The initial configuration of  $T$ reading the  infinite word $\sigma \in \Sio$ 
is  $q_0.\sigma$.  
\nl A computation of $T$ reading $\sigma \in \Sio$ is an infinite sequence of 
configurations $\alpha_0, \alpha_1, \alpha_2, \ldots ,  \alpha_i, \ldots $~~, 
where $\alpha_0=q_0.\sigma$ is the  initial configuration and for all integers $i\geq 0$, 
 $\alpha_i=u_i.q_i.v_i$ is the $(i+1)^{th}$ configuration. 
\nl The computation is successful if and only if there exists a final state $q_f \in F$ and 
infinitely many integers $i$ such that $q_i=q_f$. 

\hs We can now use a similar reasoning as in the classical 
proof of the undecidability of the emptiness 
problem for recognizable languages of finite pictures,  
 \cite[p. 34]{gr}. 
\nl We can define a set of tiles $\Delta$ in  such a way that for $\sigma \in \Sio$, 
a run $\rho$ of the tiling system 
$\mathcal{T}$=$(\Si, \Ga \cup Q, \Delta, F)$ over the infinite picture $\sigma^B$ 
satifies:
$$\mbox{ for each integer } i \geq 0 ~~~~ 
\rho(0, i).\rho(1, i).\rho(2, i)\ldots = \alpha_i=u_i.q_i.v_i$$

\noi i.e. $\rho(0, i).\rho(1, i).\rho(2, i)\ldots $ is the $(i+1)^{th}$ configuration of 
$T$ reading the $\om$-word $\sigma \in \Sio$. 
\nl Thus the B\"uchi tiling system $(\mathcal{T},$$ F)$ recognizes the language 
$L^B$.  }

\hs The following lemma is easy to prove. Details are left to the reader. 

\begin{Lem}\label{lem1'}
 Let $\alpha$ be a countable ordinal $\geq 2$. 
If $L \subseteq \Sio$ is  ${\bf \Si^0_\alpha}$-complete (respectively,  
${\bf \Pi^0_\alpha}$-complete, ${\bf \Si^1_1}$-complete),  
then  $L^B$ is ${\bf \Si^0_\alpha}$-complete (respectively,  
${\bf \Pi^0_\alpha}$-complete, ${\bf \Si^1_1}$-complete). 
\end{Lem} 

\noi In particular, for each alphabet $\Si$ having at least two letters, we get 
some ${\bf \Si^1_1}$-complete language of $\om$-pictures in the form $L^B$ because it is well 
known that there exist some ${\bf \Si^1_1}$-complete $\om$-languages $L \subseteq \Sio$ 
accepted by some B\"uchi (or Muller) Turing machine.  
\nl Notice that the ${\bf \Si^1_1}$-complete 
B\"uchi recognizable language $T_2 \subseteq\{0, 1, \$\}^{\om,\om}$ 
of all $\om$-pictures encoding an $\om$-tree with an infinite path given in \cite{atw} is  
also in that form. 

\hs  To determine the ranks of Borel languages of $\om$-pictures  we shall need 
to consider the first 
non-recursive ordinal which is called 
the Church Kleene ordinal and is usually denoted 
by $\om_1^{CK}$ ~ \cite{mos}.

\begin{Pro} Let $\Si$ be a finite alphabet having at least two letters. 
\noi
\begin{enumerate} 
\ite[(a)]
If  $L \subseteq \Si^{\om, \om}$  is 
B\"uchi recognizable by a finite tiling system and is a Borel set 
 of  rank $\alpha$, then $\alpha$   is smaller than $\om_1^{CK}$.  

\ite[(b)]  For every non null countable ordinal $\alpha < \om_1^{CK}$, there exists some  
language of infinite pictures $L \subseteq \Si^{\om, \om}$ which is 
B\"uchi recognizable by a finite tiling system and is a Borel set 
of rank $\alpha$. 
\end{enumerate}
\end{Pro}

\proof{  
\nl {\bf (a).} It was proved in \cite{atw} that every language $L$ of infinite pictures which is 
B\"uchi recognizable by a finite tiling system is definable by an existential second order 
formula of arithmetic. It is well known that this implies that $L$ is a $\Si_1^1$-set 
(lightface) and that if moreover 
$L$ is a Borel set then its Borel rank is smaller than $\om_1^{CK}$, 
see \cite{mos}.  
\nl  {\bf (b).}
For $\alpha=1$ it is well known that a ${\bf \Si^0_1}$-complete  set is simply an 
open but non closed set and that a ${\bf \Pi^0_1}$-complete  set 
is simply a closed but non open set.  
 For example $O=\{p \in \Si^{\om, \om} \mid 
\exists i \geq 1, \exists j \geq 1~~ p(i, j)=B\}$ is 
a ${\bf \Si^0_1}$-complete  subset of $\Si^{\om, \om}$, 
and $C=\{p \in \Si^{\om, \om} \mid \fa i \geq 1,  \fa j \geq 1~~ p(i, j)=B\}$ is 
a ${\bf \Pi^0_1}$-complete  subset of $\Si^{\om, \om}$.  
 These two languages are in $TS(\Si^{\om, \om})$. 
\nl  On the other hand it is  well known 
that, for every non null countable ordinal $\alpha < \om_1^{CK}$, 
 there exists some ${\bf \Si^0_\alpha}$-complete  $S_\alpha$ and   some 
${\bf \Pi^0_\alpha}$-complete  $P_\alpha$, subsets of $\Sio$, 
 which are effective, i.e. which are in the class of 
$\Si_1^1$ (lightface) subsets of $\Sio$ 
accepted by some Turing machine with a B\"uchi acceptance 
condition, \cite{mos,sta}. 
Then by Lemma \ref{lem1'}  the language 
 $(S_\alpha)^B \subseteq \Si^{\om, \om}$ (respectively,  
$(P_\alpha)^B \subseteq \Si^{\om, \om}$) is   
 ${\bf \Si^0_\alpha}$-complete (respectively,  
 ${\bf \Pi^0_\alpha}$-complete) and by  Lemma \ref{lem2'} 
these languages  are in $TS(\Si^{\om, \om})$. 
 }

\hs   In conclusion of this section, 
these results show that B\"uchi tiling systems have a much greater accepting power than 
automata  reading $\om^2$-words for acceptance of languages of infinite pictures.  
 
\section{Decision Problems}

 In a recent paper we have proved several undecidability results for 
infinitary rational relations \cite{rel-dec}. These results were deduced from  an 
extreme separation result, proved using the undecidability of the universality problem 
for finitary rational relations and the existence of a ${\bf \Si^1_1}$-complete 
infinitary rational relation stated in another paper \cite{relrat}.  
\nl We shall use in this section a very similar technique, using this time 
the undecidability of the emptiness  problem for languages of finite pictures 
and the existence of a ${\bf \Si^1_1}$-complete language of $\om$-pictures. 
In a similar way we shall see that this implies several undecidability 
results. 

\begin{Pro}\label{F} Let $\Ga=\{0, 1, \#\}$, then there exists a family $\mathcal{F}$ 
of  B\"uchi-recognizable languages of $\om$-pictures over $\Ga$, 
such that,  for $L \in \mathcal{F}$,  either 
$L=\emptyset$   or $L$  is a  ${\bf \Si^1_1}$-complete subset of $\Ga^{\om, \om}$,  
but one cannot decide which case holds. 
\end{Pro}

\proof{  We have seen that there exists a B\"uchi-recognizable 
language  $T \subseteq\{0, 1\}^{\om,\om}$ 
which is ${\bf \Si^1_1}$-complete.  
\nl  On the other side the emptiness problem for recognizable languages of finite pictures 
is known to be undecidable: if $\Si$ is an alphabet having at least one letter 
then it is undecidable 
whether  a given recognizable language $L \subseteq \Si^{\star, \star}$ is empty, see 
\cite{gr}. 

\hs  Let us define, for a finite picture $p\in \Si^{\star, \star}$ over a finite alphabet 
$\Si$ and an infinite picture 
$p'\in \Si_1^{\om,\om}$ over the alphabet  $\Si_1=\{0, 1\}$,  
the infinite picture $p \bullet p'$ over the alphabet 
$\Ga=\Si \cup \Si_1 \cup \{\#\}$. We assume that 
$\hat{\Si}=\Si \cup \{\#\}$, $\hat{\Si_1}=\Si_1  \cup \{\#\}$, and 
$\hat{\Ga}=\Ga \cup \{\#_1\}$, 
where $\#_1$ is a new letter different from the letter $\#$. 
\nl If $p$ is a finite picture of size $(m, n)$, 
the $\om$-picture $p \bullet p'$ over $\Ga$ is defined by: 
\nl $p \bullet p'(0, i)=p \bullet p'(i, 0)=\#_1$ for all integers $i \geq 0$, 
\nl $p \bullet p'(i, j)=p(i-1, j-1)$ for all integers $i \in \{1, \ldots , m+2\}$ and 
$j \in \{1, \ldots , n+2\}$, 
\nl $p \bullet p'(i, j)=\#$ for all integers $i \in \{1, \ldots , m+2\}$ and $j\geq n+2$, 
\nl $p \bullet p'(i, j)=\#$ for all integers $i\geq m+2$ and $j \in \{1, \ldots , n+2\}$,  
\nl  $p \bullet p'(i, j)=p'(i-(m+2), j-(n+2))$ for all integers $i \geq m+2$ and $j\geq n+2$. 

\hs The intuitive idea is  to construct an infinite picture $p \bullet p'$ having a prefix
$p$ ``followed" by the infinite picture $p'$, to 
``complete"  elsewhere by some letters 
$\#$ and then to border with letters $\#_1$ 
 to get an $\om$-picture in $\Ga^{\om,\om}$.  

\hs For a language $L \subseteq \Si^{\star, \star}$ we set 
$L \bullet T=\{ p\bullet p' \mid p\in L \mbox{ and } p'\in T \}.$
 It is easy to see that if $L$ is a recognizable language of finite pictures then 
$L \bullet T$ is a B\"uchi-recognizable language of $\om$-pictures because $T$ is also 
B\"uchi-recognizable.  There are now two cases:

\hs $(1)$ If $L$ is empty then 
$L \bullet T$ is empty too. 

\hs $(2)$  If $L$ is non-empty there is some 
finite picture $p\in L \subseteq \Si^{\star, \star}$. 
Let then $\psi_p$ be the mapping from 
$\Si_1^{\om,\om}$ into 
$(\hat{\Si} \cup\Si_1)^{\om,\om}$ defined by $\psi_p(p')=p\bullet p'$. 
\nl It is easy to see that the mapping $\psi_p$ is continuous and that 
$\psi_p^{-1}(L \bullet T)= T$. But  $T$ is ${\bf \Si^1_1}$-complete and 
$L \bullet T$,  as well as 
every B\"uchi-recognizable language of infinite pictures, is a ${\bf \Si^1_1}$-set
because it is definable by an existential second order monadic formula,  \cite{atw}. 
This implies that $L \bullet T$ is a ${\bf \Si^1_1}$-complete set.

\hs We can now choose the family $\mathcal{F}$ to be  the family of languages $L \bullet T$ 
obtained with $\Si=\{0\}$ and $L$ running over recognizable languages of pictures over 
$\Si$.    }  

\hs In order to disprove the existence of decision procedures which test B\"uchi-recognizable 
$\om$-picture languages for E-, respectively A-recognizability, we shall need the following 
lemmas. 

\begin{Lem}\label{lem1}
Let $\Si$ be an alphabet having at least two letters and $L \subseteq \Si^{\om, \om}$ be a 
E-recognized language of $\om$-pictures. Then $L$ is a ${\bf \Si_{2}^0}$-subset of 
$\Si^{\om, \om}$.   
\end{Lem}

\proof{  Let $\Si$ be an alphabet having at least two letters and 
$L \subseteq \Si^{\om, \om}$ be a language of $\om$-pictures which is 
E-recognized by $(\mathcal{A},$$ F)$, where $\mathcal{A}$=$(Q, \Si, \Delta)$ 
is a tiling system 
and $F \subseteq Q$. 

\hs Let $R=\{(p, \rho) \in \Si^{\om, \om} \times Q^{\om^2} 
\mid \rho \mbox{ is a run of } \mathcal{A} \mbox{ on }$ $ p \}$. It is easy to see that 
$R$ is a closed subset of $\Si^{\om, \om} \times Q^{\om^2}$ where 
$\Si^{\om, \om} \times Q^{\om^2}$ is equipped with the classical product topology. 
\nl  Let $R_E=\{(p, \rho) \in \Si^{\om, \om} \times Q^{\om^2} 
\mid \exists v \in \om^2 ~~ \rho(v) \in F  \}$. It is easy to see that $R_E$ is an open subset 
of $\Si^{\om, \om} \times Q^{\om^2}$. 

\hs Then the set $R \cap R_E$ is a boolean combination of open sets. In particular 
 it is a ${\bf \Si_{2}^0}$-subset of $\Si^{\om, \om} \times Q^{\om^2}$, i.e. a countable union 
of closed subsets of $\Si^{\om, \om} \times Q^{\om^2}$. 
\nl But the topological space $\Si^{\om, \om} \times Q^{\om^2}$ is compact because it is the
 product of two compact spaces. Thus every closed subset of $\Si^{\om, \om} \times Q^{\om^2}$ 
is also compact. Therefore $R \cap R_E$ is a  countable union of compact subsets of 
 $\Si^{\om, \om} \times Q^{\om^2}$. 

\hs But the language $L$ is E-recognized by  $(\mathcal{A},$$ F)$ so it
is the projection of the set $R \cap R_E$ onto $\Si^{\om, \om}$.  
The projection from $\Si^{\om, \om} \times Q^{\om^2}$ onto $\Si^{\om, \om}$ 
is continuous and the continuous image of a compact set is a compact set. Thus the  
language $L$ is a countable union of compact sets hence  it
is a countable union of closed  sets, 
i.e. a  ${\bf \Si_{2}^0}$-subset of 
$\Si^{\om, \om}$.   
 } 

\begin{Lem}\label{lem2}
Let $\Si$ be an alphabet having at least two letters and $L \subseteq \Si^{\om, \om}$ be a 
A-recognized language of $\om$-pictures. Then $L$ is a closed  subset of 
$\Si^{\om, \om}$.   
\end{Lem}

\proof{  Let $\Si$ be an alphabet having at least two letters and 
$L \subseteq \Si^{\om, \om}$ be a language of $\om$-pictures which is 
A-recognized by $(\mathcal{A},$$ F)$, where 
 $\mathcal{A}$=$(Q, \Si, \Delta)$ is a tiling system and $F \subseteq Q$. 

\hs We call $cl(L)$ the topological closure of $L$ and we are going to prove that 
$L=cl(L)$. For that purpose consider an $\om$-picture $p$ in $cl(L)$. For all integers 
$i\geq 1$ there is some $\om$-picture $p_i \in L$ such that 
$p$ $|$ $_{ \{0, 1, \ldots, i\} \times \{0, 1, \ldots, i\} }$=
$p_i$  $|$$_{  \{0, 1, \ldots, i\} \times \{0, 1, \ldots, i\} }$  

\hs For each integer $i\geq 1$, $p_i \in L$ thus there is some run $\rho_i$ of $\mathcal{A}$ 
on $p_i$ such that for all $v\in \om^2$ ~ $\rho_i(v)\in F$. 
\nl  Consider now the partial runs 
$\rho'_{i,j}=\rho_i$ $|$ $_{  \{0, 1, \ldots, j\} \times \{0, 1, \ldots, j\} }$, 
for $j\leq i$,  of 
$\mathcal{A}$ on  the restriction  of $p_i$ (hence also of $p$) to 
$\{0, 1, \ldots, j\} \times \{0, 1, \ldots, j\}$. 

\hs We can now reason as in the proof of Theorem 4 (a) in \cite{atw}. 
These partial runs $\rho'_{i,j}$ are 
arranged in a finitely branching tree, via the extension relation. 
By construction  this tree is infinite so by K\"{o}nig's Lemma there is an infinite path. 
This infinite path determines a run of $\mathcal{A}$ on $p$ which is A-accepting  
thus $p\in L$.

\hs We have then proved that $cl(L)\subseteq L$ so $L=cl(L)$ and $L$ is a closed subset of 
$\Si^{\om, \om}$.   
 }

\hs We can now infer the following result. 

\begin{Pro} 
There are no  decision procedures which test B\"uchi-recognizable 
$\om$-picture languages for E-, respectively A-recognizability. 
\end{Pro}

\proof{  Consider the family $\mathcal{F}$ 
of $\om$-picture B\"uchi-recognizable languages  over $\Ga$, 
such that,  for $L \in \mathcal{F}$,  either 
$L=\emptyset$   or $L$  is a  ${\bf \Si^1_1}$-complete subset of $\Ga^{\om, \om}$. 

\hs In the first case, $L$ is obviously A-recognizable and E-recognizable. 
 In the second case $L$  is  ${\bf \Si^1_1}$-complete so in particular it is not a Borel 
subset of $\Ga^{\om, \om}$. By Lemmas \ref{lem1} and \ref{lem2} it cannot be E-recognizable 
(respectively, A-recognizable). 
 But   one cannot decide which case holds.  
 }

\hs As remarked in \cite{atw} Staiger-Wagner  and co-B\"uchi 
recognizability reduces to E-recognizability so the above proof can be applied 
to Staiger-Wagner  and co-B\"uchi 
recognizability instead of E-recognizability.

\hs Proposition \ref{F} gives an extreme separation result which implies other undecidability 
results. For example for any Borel class ${\bf \Si_{\alpha}^0}$ or   ${\bf \Pi_{\alpha}^0}$, 
$\alpha$ being  a countable ordinal $\geq 1$,  it 
is undecidable whether a given B\"uchi-recognizable language of $\om$-pictures is in 
${\bf \Si_{\alpha}^0}$ (respectively  ${\bf \Pi_{\alpha}^0}$). It is even undecidable 
whether a given B\"uchi-recognizable language   of $\om$-pictures is a Borel set or a 
${\bf \Si^1_1}$-complete set. 
\nl Remark that the same result holds if we replace Borel classes by arithmetical classes 
$\Si_i$ or $\Pi_i$, $i\geq 1$, and the class of Borel sets by the class of arithmetical sets 
$\cup_{n\geq 1} \Si_n = \cup_{n\geq 1} \Pi_n$.
\nl  These results show a great contrast with the case of recognizable 
languages of infinite words 
where such problems are decidable \cite{la}. 

\hs Recall now the following definition, see \cite{atw}: 
 a tiling system is called deterministic if on any picture 
it allows at most one tile covering the origin, the state assigned to position 
$(i+1, j+1)$ is uniquely determined by the states at positions $(i, j), (i+1, j), (i, j+1)$ 
and the states at the border positions $(0, j+1)$ and $(i+1, 0)$ are determined by the state 
$(0, j)$, respectively $(i, 0)$. 
\nl As remarked in \cite{atw}, the hierarchy proofs of the classical 
Landweber hierarchy defined using deterministic $\om$-automata ``carry over without essential 
changes to pictures". In particular it is easy to see that a language of $\om$-pictures which 
is B\"uchi-recognized by a \de  tiling system is a ${\bf \Pi^0_2}$-set. 
\nl Remark that if we use the classical Muller acceptance 
condition instead of the B\"uchi condition, we can easily show, 
as in the case of infinite words, that a language of $\om$-pictures which 
is Muller-recognized by a \de  tiling system is a boolean combination of ${\bf \Pi^0_2}$-sets.

\hs We now state the following results. 
 
\begin{Pro} Let $\Ga=\{0, 1, \#\}$ as in Proposition \ref{F}.  It 
is undecidable for  a given B\"uchi-recognizable language $L \subseteq \Ga^{\om, \om}$ whether:  
\begin{enumerate}
\ite[(1)] $L$ is B\"uchi-recognized by a \de  tiling system. 
\ite[(2)] $L$ is Muller-recognized by a \de  tiling system. 
\ite[(3)] its complement  $\Ga^{\om, \om}-L$ is B\"uchi-recognizable. 
\ite[(4)] $\bar{L}$ is $\om^2$-regular. 
\end{enumerate}
\end{Pro}

\proof{  Consider the family $\mathcal{F}$ 
of B\"uchi-recognizable  $\om$-picture languages given by Proposition \ref{F}. 
Then two cases may happen for $L \in \mathcal{F}$: either 
$L$ is empty or $L$ is ${\bf \Si^1_1}$-complete. 
\nl   In the first case $L$ is obviously recognized  by a \de B\"uchi or Muller
tiling system;  its complement  $\Ga^{\om, \om}-L=\Ga^{\om, \om}$ is B\"uchi-recognizable and 
$\bar{L}$ is $\om^2$-regular. 
\nl   In the second case $L$ is ${\bf \Si^1_1}$-complete.  Thus $L$ is not a Borel set hence it 
is neither  B\"uchi nor  Muller-recognized by any  \de  tiling system 
and  $\bar{L}$ is not $\om^2$-regular. 
 Moreover in this second case its complement $\Ga^{\om, \om}-L$ is  
 a ${\bf \Pi^1_1}$-complete subset of  $\Ga^{\om, \om}$.    It is well known that a 
${\bf \Pi^1_1}$-complete set is not a ${\bf \Si^1_1}$-set thus it cannot be 
B\"uchi-recognizable. 
\nl But Proposition \ref{F} states that one cannot decide which case holds.   
 } 

\nonumsection{Acknowledgements}
 Thanks to  the anonymous referee 
for useful comments on a preliminary version of this paper.

\newpage
\nonumsection{References}

\newpage 

\begin{center}
{\bf ANNEXE : \\
ERRATUM TO THE PAPER : 
\\ \hs ON RECOGNIZABLE LANGUAGES\\
        OF INFINITE PICTURES}
\end{center}

\hs Recall  first  that  the following result was stated as  Proposition 4.7 in \cite{Finkel04}. 

\hs {\bf  Proposition 4.7 }   {\it 
Let $\Si$ be a finite alphabet having at least two letters. 
\noi
\begin{enumerate} 
\ite[(a)]
If  $L \subseteq \Si^{\om, \om}$  is 
B\"uchi recognizable by a finite tiling system and is a Borel set 
 of  rank $\alpha$, then $\alpha$   is smaller than $\om_1^{\mathrm{CK}}$.  

\ite[(b)]  For every non null countable ordinal $\alpha < \om_1^{\mathrm{CK}}$, there exists some  
language of infinite pictures $L \subseteq \Si^{\om, \om}$ which is 
B\"uchi recognizable by a finite tiling system and is a Borel set 
of rank $\alpha$. 
\end{enumerate}
}

\noi Item (a) of this result was deduced from the fact  that if $L$ is a (lightface) $\Si_1^1$-set 
 and that if moreover 
$L$ is a Borel set then its Borel rank is smaller than $\om_1^{\mathrm{CK}}$.  
 This fact,  which is true if we replace the (lightface)  class  $\Si_1^1$ by the (lightface)  class $\Delta_1^1$,  is actually not true
 and the given  reference [Mos80] does not contain this result. 

\hs   
Kechris, Marker and Sami proved in \cite{KMS89} that the supremum 
of the set of Borel ranks of  (lightface) $\Pi_1^1$, so also of  (lightface) $\Si_1^1$,  sets is the ordinal $\gamma_2^1$. 
\nl This ordinal is precisely defined in \cite{KMS89}.  Kechris, Marker and Sami proved that the ordinal $\gamma_2^1$
 is strictly greater than the ordinal $\delta_2^1$ which is the first non $\Delta_2^1$ ordinal. 
Thus in particular it holds that $ \om_1^{\mathrm{CK}} < \gamma_2^1$.  
Notice that the exact value of the ordinal $\gamma_2^1$ may depend on axioms of  set theory.  
For more details,  the reader is  referred to  \cite{KMS89} and to 
a textbook of set theory like \cite{Jech}. 

\hs Notice  that it seems still unknown  whether {\it every } non null ordinal $\gamma < \gamma_2^1$ is the Borel rank 
of a (lightface) $\Pi_1^1$ (or $\Si_1^1$) set. 
On the other hand,  for every non null  ordinal $\alpha < \om_1^{\mathrm{CK}}$, there exist some  
 ${\bf \Si}^0_\alpha$-complete and some  ${\bf \Pi}^0_\alpha$-complete sets in the class $\Delta_1^1 \subset \Si_1^1$. 
This is a well  known fact of Effective Descriptive Set Theory which  is proved in detail in \cite{Fink-Lec2}. 

\hs We can now state the following result which corrects the above false Proposition 4.7. 

\hs {\bf  Theorem}   {\it 
\noi 
\begin{enumerate} 
\ite[(a)]  The Borel hierarchy of the class $\mathcal{C}$ of B\"uchi 
recognizable language of infinite pictures is equal to the Borel hierarchy of the class $\Sigma^1_1$. 
\ite[(b)]   $\gamma_2^1=Sup ~~ \{ \alpha \mid \exists L \in \mathcal{C} \mbox{ such that }$$L $$\mbox{ is a Borel set of rank } \alpha \}.$  
\ite[(c)]  For every non null  ordinal $\alpha < \om_1^{\mathrm{CK}}$, 
there exists some  
${\bf \Si}^0_\alpha$-complete and some ${\bf \Pi}^0_\alpha$-complete
$\om$-languages in the class $\mathcal{C}$.  
\end{enumerate}
}

\noi 
This result follows easily from the proof of 
Lemmas 4.5 and 4.6 of \cite{Finkel04} and from the above cited result of Kechris, Marker and Sami proved in \cite{KMS89}. 

\hs Notice that a very similar result was obtained in \cite{Fin-mscs06}
 for the class of $\om$-languages accepted by (real time) one counter B\"uchi automata, and in \cite{Fink-Wd} for the class of infinitary rational relations accepted by 
$2$-tape  B\"uchi automata. 

\nonumsection{References}

\end{document}